

\input phyzzx
\font\first=cmb10 scaled\magstep3
\font\second=cmr10 scaled\magstep2
{\nopagenumbers
\rightline{Preprint RIMS 1003}
\rightline{December 1994}
\rightline{(Revised, January 16)}
\rightline{hep-th/9412237}
\vskip2cm
\centerline{\first Generalized Quantum Inverse Scattering}
\vskip3cm
\centerline{{{\second Christian Schwiebert}\footnote{\star}
{Supported by the Science and Technology Fellowship Programme for
Japan under the auspices of the Commission of the European
Communities}}\footnote{\diamond}{e-mail:
{\tt cbs@kurims.kyoto-u.ac.jp}}}
\centerline{\it Research Institute for Mathematical Sciences}
\centerline{\it Kyoto University, Sakyo-ku, 606 Kyoto, Japan}
\vskip5cm
\centerline{ABSTRACT}
\smallskip
\noindent
A generalization of the quantum inverse scattering method is proposed
replacing the quantum group $RLL$ commutation relations of Lax
operators by reflection equation type $RLRL$ commutation relations.
Under some natural assumptions the most general algebra of this type
allowing to construct the neccessary integrals of motion is found. It
serves to describe Lax operators with completely non-ultralocal
commutation relations. An example of this new formalism is an
integrable model on monodromies of flat connections on a Riemann
surface which is related to the XXZ quantum spin chain.
\vfill
\break}
\pageno=2


\def \Ale {A. Yu. Alekseev, Uppsala Univ. preprint, (hep-th/9311074)}
\def \AFS {A. Yu. Alekseev, L. D. Faddeev and M. A.
Semenov-Tian-Shansky, Commun. Math. Phys. {\bf 149}, 335 (1992)}
\def \AGS {A. Yu. Alekseev, H. Grosse and V. Schomerus, Harvard Univ.
preprints HUTMP-94-B336, HUTMP-94-B337}
\def \Che {I. V. Cherednik, Teor. Mat. Fiz. {\bf 61}, 55 (1984)}
\def \Dri {Drinfel'd, \ in: Proc. ICM-86, ed. A. M. Gleason, AMS,
Providence 1987}
\def \Fad {L. D. Faddeev, Stony Brook Univ. preprint ITP-SB-94-11}
\def \FRT {L. D. Faddeev, N. Yu. Reshetikhin and L. A. Takhtajan,
Alg. i Anal. {\bf 1}, 178 (1989)\ \lbrack \thinspace in Russian,
English transl.: Leningrad Math. J. {\bf 1}, 193 (1990) \rbrack}
\def \Hla {L. Hlavaty, Prague Univ. preprint, (hep-th/9412142)}
\def \HlKu {L. Hlavaty and A. Kundu, Bonn Univ. preprint,
(hep-th/9406215)}
\def \KaZa {M. Karowski and A. Zapletal, Nucl. Phys. {\bf B419}, 567
(1994)}
\def \Kor {V. E. Korepin, Zap. Nauch. Sem. LOMI {\bf 101}, 90 (1981)
\lbrack \thinspace in Russian, Engl. transl.: J. Sov. Math. {\bf 23},
2429 (1983) \rbrack}
\def \KSS {P. P. Kulish, R. Sasaki and C. Schwiebert, J. Math. Phys.
{\bf 34}, 286 (1993)}
\def \Kul {P. P. Kulish, Teor. Mat. Fiz. {\bf 94}, 137 (1993)}
\def \KuSa {P. P. Kulish and R. Sasaki, Prog. Theor. Phys. {\bf 89},
741 (1993)}
\def \KuSk {P. P. Kulish and E. K. Sklyanin, J. Phys. {\bf A25}, 5963
(1992)}
\def \Mad {S. Majid, J. Math. Phys. {\bf 32}, 3246 (1991)}
\def \FrMa {L. Freidel and J. M. Maillet, Phys. Lett. {\bf 262B},
278 (1991); {\bf 263B}, 403 (1991)}
\def \Sch {C. Schwiebert, J. Math. Phys. {\bf 35}, 5288 (1994)}
\def \Skl {E. K. Sklyanin, J. Phys. {\bf A21}, 2375 (1988)}
\def \STF {E. K. Sklyanin, L. A. Takhtajan and L. D. Faddeev, Teor.
Mat. Fiz. {\bf40}, 194 (1979)}
\def \Tak {L. A. Takhtajan, \ in: \lq\lq Introduction to quantum
groups and integrable models of quan- tum field theory\rq\rq,
eds. M. L. Ge and B. H. Zhao, World Scientific, Singapore 1991}


\def\sl#1{$sl_q(#1)$}
\def\Rt{\widetilde R}
\def\St{\widetilde S}
\def\Ri{R^{-1}}
\def\r#1{R^{#1}(\l,\m)}
\def\rt#1{\Rt^{#1}(\m,\l)}
\def\K#1#2{K^{#1}_{#2}}
\def\L#1#2{L^{#1}_{#2}}
\def\S#1#2{S^{#1}_{#2}}
\def\A{{\cal A}}
\def\one{{\bf 1}}
\def\tr{{\it Tr}}
\def\a{\alpha}
\def\b{\beta}
\def\c{\gamma}
\def\e{\varepsilon}
\def\l{\lambda}
\def\m{\mu}
\def\w{\omega}
\def\NPrefs{\let\refmark=\NPrefmark}
\NPrefs


\chapter{INTRODUCTION}

The quantum inverse scattering method (QISM) was developed at the end
of the '70s by the Leningrad School in order to handle quantum
versions of classically integrable models in 1+1 dimensions
\Ref\rSTF{\STF}. The algebraic framework that characterizes the method
was interpreted later as a deformation of Poisson-Lie groups
\Ref\rDri{\Dri} and led to the discovery of quantum groups (QG).

For our purposes the formulation on discretized space with $N$ lattice
points (time dependence is suppressed throughout) will be appropriate.
Given a model on such a space it is sometimes possible to find
so-called Lax operators satisfying a linear differential equation
encoding the equations of motion which can be derived from the
Hamiltonian of the model. Then the fundamental commutation relations
of the Lax operators $L^n(\l)$ at lattice sites $n = 1,\ldots,N$ have
the form
$$
\eqalign{R_{12}(\l-\m) \L{n}1(\l) \L{n}2(\m) &= \L{n}2(\m) \L{n}1(\l)
R_{12}(\l-\m) \cr
\L{m}1(\l) \L{n}2(\m) &= \L{n}2(\m) \L{m}1(\l) , \qquad m \not= n}
\eqn \fcr
$$
where all quantities depend on spectral parameters $\l,\m \in {\cal
\bf C}$. Lax operators of different sites commute, this is referred to
as ultralocality. Indices $i = 1,2$ denote the auxiliary spaces $V_i$
of dimension $r$ on which the matrices $\L{n}1 = L^n \otimes \one,
\L{n}2 = \one \otimes L^n$ and $R_{12}$ act nontrivially (we employ
the usual quantum group terminology, see \REF\rFRT{\FRT}
\REF\rTak{\Tak} \refmark{\rFRT,\rTak} for example). The matrix $R$
of dimension $2r \times 2r$ satisfies the Yang-Baxter equation (YBE)
$$
R_{12}(\l-\m) R_{13}(\l) R_{23}(\m) = R_{23}(\m) R_{13}(\l)
R_{12}(\l-\m) . \eqn \ybe
$$
As a consequence of ultralocality the product $L^{n+1}(\l) L^{n}(\l)$
also satisfies the fundamental commutation relations (this is
equivalent to existence of a coproduct for the algebra of Lax
operators). By iteration the same holds for the monodromy matrix
$$
T(\l) = L^N(\l) L^{N-1}(\l) \cdots L^1(\l) , \eqn \mon
$$
and \fcr\ implies that the operators $t(\l) = \tr\lbrack T(\l)
\rbrack$ are commuting for different values of spectral parameters
$$
\lbrack t(\l),t(\m)\rbrack = 0 . \eqn \zero
$$
Expanding $t(\l)$ in powers of $\l$ one obtains a set of mutually
commuting operators and, moreover, they are conserved because it can
be shown that the Hamiltonian is among them. This gives a systematic
procedure to obtain all the integrals of motion neccessary for
integrability of the model.

In this paper we suggest a generalization of the quantum inverse
scattering method based on reflection equation (RE) type algebras
rather than the $RLL$ algebra. This way we can describe Lax operators
which are mutually non-commuting for any pair of lattice sites. The
idea for such a generalized formalism relies on our experience with
so-called extended RE algebras \Ref\rSch{\Sch} (see also
\REF\rKul{\Kul} \REF\rKuSa{\KuSa} \refmark{\rKul,\rKuSa}) and the
explicit example of an integrable model that can be constructed out of
them \Ref\rAle{\Ale}. Motivated by Chern-Simons theory in
\refmark{\rAle} an integrable model was introduced on the moduli space
of flat connections on Riemann surfaces which is related to the $XXZ$
quantum spin chain. For motivation and to show that the subsequently
developed formalism is not empty we outline the main points of that
model in the following.  We use the notation of \refmark{\rSch} where
part of the construction was also carried out in order to describe the
braid group on a handlebody.

Single reflection equations without spectral parameters were
investigated in \REF\rKSS{\KSS} \REF\rKuSk{\KuSk}
\refmark{\rKSS,\rKuSk}. Extended RE algebras consist of $N$ reflection
equations for matrices $K^n, n=1,\ldots,N$ of size $r \times r$ with
operator valued entries belonging to some algebra \A, and additional
commutation relations between them
$$
\eqalign{R K^n_1 \Rt K^n_2 &= K^n_2 R K^n_1 \Rt  \cr
R K^m_1 R^{-1}K^n_2 &= K^n_2 R K^m_1 R^{-1} , \qquad m>n} \eqn \re
$$
where $R = R_{12},\ \Rt = R_{21} \equiv P R P$ and $P$ is the
permutation operator (we suppress indices where possible). Here the
matrix $R$ of size $2r \times 2r$ is any invertible solution of the
constant (spectral parameter independent) Yang-Baxter equation \ybe,
but for the following example we restrict it to the $R$-matrix of \sl2
given by
$$
R = \pmatrix{q&0&0&0 \cr 0&1&0&0 \cr 0&\w&1&0 \cr
0&0&0&q \cr} , \quad \w=q-q^{-1} .  \eqn\rmat
$$
One of the main properties of \re\ is invariance w.r.t. QG coaction,
i.e.  the transformed $K^n_T = T^{(n)} K^n (T^{(n)})^{-1}$ are also
solutions of \re\ if $\K{n}1 T^{(m)}_2 = T^{(m)}_2 \K{n}1$ for all
$m,n$, i.e. all elements of $K^n$ and $T^{(m)}$ are commuting and
$T^{(n)}$ obeys the system of QG relations
$$
\eqalign{R\, T^{(n)}_1 T^{(n)}_2 &= T^{(n)}_2 T^{(n)}_1 R  \cr
R\, T^{(m)}_1 T^{(n)}_2 &= T^{(n)}_2 T^{(m)}_1 R , \qquad m>n.}
\eqn \rtt
$$
A further property of \re\ is that $K^{n+1} K^n$ satisfies also the
extended RE algebra (this can be interpreted as a braided coproduct
for the extended RE algebra in the sense of \Ref\rMad{\Mad}). In fact,
the same holds for any strictly ordered product of $K$-matrices with
values of indices decreasing from left to right, for the second
equation of \re\ it means that $m$ has to be greater than the largest
index in the ordered product.

Now we describe a representation of \re\ in terms of \sl2 algebra
generators $H,X^{\pm}$ which obey the relations
$$
\lbrack H,X^{\pm}\rbrack = \pm 2 X^{\pm} , \qquad
\lbrack X^+,X^- \rbrack = \w^{-1} (q^H - q^{-H}) . \eqn \sltwo
$$
They can be conveniently rewritten in terms of three matrix equations
$$
\Rt L^{\e_1}_1 L^{\e_2}_2 = L^{\e_2}_2 L^{\e_1}_1 \Rt , \qquad
(\e_1,\e_2) \in  \{(+,+),(+,-),(-,-)\}  \eqn \rll
$$
where the $L^{\pm}$ are triangular matrices expressed in terms of \sl2
generators
$$
L^+ = \pmatrix{q^{H/2} & q^{-1/2} \w X^- \cr 0 & q^{-H/2} \cr} ,
\qquad L^- = \pmatrix{q^{-H/2} & 0 \cr -q^{1/2} \w X^+ & q^{H/2} \cr}.
\eqn \lplm
$$
These matrices can be formally inverted by applying the so-called
antipode map $S$ to the generators defined by $ S(H) = - H,\
S(X^{\pm}) = - q^{\mp 1} X^{\pm} $. Define
$$
K^1 = S(L^-) L^+ =  \pmatrix{q^H & q^{-1/2} \w q^{H/2} X^- \cr
q^{-1/2} \w X^+ q^{H/2} & q^{-H} + q^{-1} \w^2 X^+ X^- \cr} ,
\eqn \lsol
$$
it is straightforward to show with help of \rll\ that $K^1$ satisfies
the first equation of \re. To represent the whole algebra \re\ we
define further
$$
\L{\pm}i = \one {\dot \otimes} \cdots \one {\dot \otimes}
L^{\pm} {\dot \otimes} \one \cdots {\dot \otimes} \one , \qquad
1 \leq i \leq N  \eqn \lipm
$$
with $L^{\pm}$ inserted into the $i$-th position. The dot over the
tensor product means matrix multiplication of these $2 \times 2$
matrices such that \lipm\ again is a $2 \times 2$ matrix whose entries
take value in the $N$-fold tensor product of the (universal
enveloping) quantum algebra \sl2. Operators contained in different
spaces of the tensor product are commuting. Setting $K_i = \S-i \L+i$,
we then have the following set of operators $K^n$ satisfying \re
$$
\K{n}{} = \S-1 \cdots \S-{n-1} K^{\phantom{-}}_n \L-{n-1} \cdots
\L-1 , \eqn \reop
$$
where we have put $\S-i \equiv S(\L-i)$ for brevity. More explicitly,
the operators defined in \reop\ are written as
$\K1{} = \K{}1,\  \K2{} = \S-1 \K{}2 \L-1,\  \ldots,\  \K{N}{} = \S-1
\cdots \S-{N-1} \K{}N \L-{N-1} \cdots \L-1$. In \reop\ we could have
equivalently used $L^+$ instead of $L^-$, see \refmark{\rSch}.

These structures can be utilized to obtain an integrable model from
monodromies of flat connections along fundamental cycles of a Riemann
surface which obey \re. Introduce the spectral parameter dependent
`Lax operators'
$$
L^n(\l) = K^n + \l \one , \eqn \lax
$$
as well as a spectral parameter dependent matrix $S(\l,\m)$ which
satisfies the YBE
$$
S(\l,\m) = \l {\Rt}^{-1} - \m R , \qquad
\St(\l,\m) = \l \Ri - \m \Rt , \eqn \smat
$$
then it can be proven with help of \re\ that these quantities satisfy
the system of equations
$$
\eqalign{S(\l,\m) \L{n}1(\l) \Rt \L{n}2(\m) &= \L{n}2(\m) R \L{n}1(\l)
\St(\l,\m) \cr  R \L{m}1(\l) \Ri \L{n}2(\m) &= \L{n}2(\m) R \L{m}1(\l)
\Ri , \qquad m>n .} \eqn \spre
$$
In order to show this the Hecke relation $P R - (P R)^{-1} = \w I$
must be used which restricts the model to the fundamental
representation of \sl2. Equations \spre\ are a spectral parameter
dependent version of the extended RE algebra but some $R$-matrices
remain constant.  Comparing them to \fcr\ we observe that they are
completely non-ultralocal.  From \re\ they inherit the property that
ordered products of Lax operators do also satisfy \spre\ such that the
monodromy $T(\l)$ defined as in \mon\ has the commutation relation
$$
S(\l,\m) T_1(\l) \Rt\, T_2(\m) = T_2(\m) R\, T_1(\l) \St(\l,\m) . \eqn
\tcom
$$
It can be proven that the quantum trace of $T(\l)$
$$
t(\l) = \tr_q \lbrack T(\l) \rbrack = \tr \lbrack M T(\l) \rbrack ,
\qquad M = \pmatrix{q^{-1} & 0 \cr 0 & q} \eqn \qutr
$$
commutes with itself for different values of spectral parameters as
in \zero, and thereby produces a family of conserved operators in
involution. It is obvious that the model can be generalized to \sl3,
etc. We refer to \REF\rAGS{\AGS} \refmark{\rAle,\rAGS} for its
relation to Chern-Simons theory. Here we only note that it
is related to the $XXZ$ quantum spin chain, however by a nonunitary
transformation. Define $H^n = \L-{n} \L-{n-1} \cdots \L-1,\
H^0 = \one$, then the transformed $L^n(\l)$
$$
H^n L^n(\l) S(H^{n-1}) = \L+n + \l \L-n , \eqn \trans
$$
is equivalent to the Lax operator of the $XXZ$ chain (for a recent
review see \Ref\eFad{\Fad}). It seems to us that the quantum group
invariant $n$-state vertex model on a torus constructed in
\Ref\rKaZa{\KaZa} is the statistical mechanics analogue of the model
above, and that \spre\ with all $R$-matrices spectral parameter
dependent should hold for the monodromies defined there (before
letting a certain parameter go to infinity). We will show in the next
section how the model fits into the general framework for theories
with mutually non-commuting Lax operators developed there.


\chapter{PROPERTIES OF GENERALIZED QISM}

The analogy of QISM based on fundamental commutation relations \fcr\
with the integrable model associated to the spectral parameter
dependent extended RE algebra \spre\ suggests to investigate the
possibility to establish a general theory of integrable models whose
Lax operators obey RE type algebras. The problem to be solved then is
what does the most general form of such an algebra look like under the
condition that a system of Hamiltonians in involution can be
constructed? We will see that the answer is unique given some
reasonable assumptions.

Generalized QISM will be based on the general spectral parameter
dependent RE
$$
\r1 \L{n}1(\l) \r2 \L{n}2(\m) = \L{n}2(\m) \r3 \L{n}1(\l) \r4 ,
\eqn \genre
$$
with, as yet, four arbitrary $R$-matrices $R^i,\ i = 1,\ldots,4$ which
have to be subjected to some consistency conditions to be derived. It
is not clear a priori what assumptions are possible such that a
consistent formalism is guaranteed. Conventional QISM was developed by
generalizing common properties of a number of models, as they are
scarce in our case we will copy properties of conventional QISM as far
as possible and be guided by the example above to arrive at a set of
natural and hopefully minimal assumptions.

We remark that a reflection equation of the type \genre\ was studied
systematically for the first time in \Ref\rChe{\Che} where it emerged
as a consistency condition for scattering off the endpoint of
particles moving on a half-line, much like the YBE is a consistency
condition for particle scattering on a line. But it appeared even
earlier in \Ref\rKor{\Kor} where it described the commutation relation
of the monodromy of Lax operators for the non-abelian quantum Toda
chain (two of the $R$-matrices there are constant like in \tcom). It
then reappeared in \Ref\rSkl{\Skl} where it was used to encode
boundary conditions for non-periodic integrable models. The fact that
there exist so-called non-ultralocal integrable models like the Toda
chain where Lax operators of neighbouring sites do not commute, but
commute for $\mid m - n \mid \geq 2$ (no commutation relations of
those Lax operators are of RE type) and nevertheless have a monodromy
obeying the RE led to a systematic study of quadratic algebras defined
by the RE in \Ref\rFrMa{\FrMa}. This will save us some work because
our monodromy matrix will also satisfy the RE which allows to take
over several results of \refmark{\rFrMa}, the following can be viewed
as a generalization of that work. Closest to our program comes
\Ref\rHlKu{\HlKu} where some constant quadratic algebra was
Yang-Baxterized to describe the non-ultralocal models mentioned above.
This gave a RE type algebra part of whose $R$-matrices then were
restricted to the identity in order to classify some known integrable
models with nearest neighbour interactions. We are not aiming at these
models but rather completely non-ultralocal models like the one
discussed above and try to find the most general form compatible with
integrability.

{\bf General construction.} The first, natural assumption is simply to
demand that the product $L^{n+1}(\l) L^n(\l)$ again satisfies RE
\genre. Upon insertion of this product into \genre\ it is easy to see
that we need a relation interchanging $\L{n}1(\l)$ and $\L{n+1}2(\m)$
of the form
$$
\L{n}1(\l) \r2 \L{n+1}2(\m) = \r\a \L{n+1}2(\m) \r\b \L{n}1(\l) \r\c ,
\eqn \inter
$$
with three more $R$-matrices $R^\a,R^\b,R^\c$ to be determined. Then
we have to use \genre\ for both $L^n$ and $L^{n+1}$ to see what
conditions on the unknown $R$-matrices arise. This analysis gets more
involved by the fact that \genre\ can be written equivalently as
$$
(\rt1)^{-1} \L{n}1(\l) \rt3 \L{n}2(\m) = \L{n}2(\m) \rt2 \L{n}1(\l)
(\rt4)^{-1} , \eqn \greequ
$$
where $\rt{i} = P \r{i} P$ is obtained by interchanging $\l
\leftrightarrow \m$ and conjugating with the permutation operator.
Hence, four cases must be considered and the $R$-matrices be fixed
differently in each case. We do not go into details here and
present the result which is surprisingly simple. There is only one
general case depending on three $R$-matrices which we choose to be
$R^1,R^2,R^\b$ and all other possibilities are contained in this
case given by
$$
\eqalign{\r1 \L{n}1(\l) \r2 \L{n}2(\m) &= \L{n}2(\m) \rt2 \L{n}1(\l)
\bigl \lbrack \rt\b \r1 (\r\b)^{-1} \bigr \rbrack \cr
\rt2 \L{n+1}1(\l) \rt\b \L{n}2(\m) &= \L{n}2(\m) \rt2 \L{n+1}1(\l)
(\rt2)^{-1} .} \eqn \nnre
$$
As in the conventional formalism we demand that the monodromy
satisfies the same equation as the individual Lax operator, this is
our second assumption. For this it is clearly neccessary that the
product $L^{n+2}(\l) L^{n+1}(\l) L^n(\l)$ and any ordered product of
Lax operators also satisfies the first equation of \nnre. We encounter
a problem then because we would need equations of the type \inter\ for
any pair of Lax operators introducing more and more unknown
$R$-matrices. The analysis for arbitrary $N$ soon becomes so
complicated that we did not attempt to find general rules which
determine the $R$-matrices that remain free (moreover an integrable
system where almost all Lax operators commute differently is hard to
conceive). The way out is to impose that pairs of Lax operators
separated by equal distances on the lattice share the same commutation
relation, this is our third assumption. It fixes the solution
uniquely, we find that Lax operators which are not on neighbouring
sites have all the same commutation relation depending only on $\r2$.
We rewrite \nnre\ in a more systematic way (renaming $\r\b
\rightarrow (\r3)^{-1}$), together with the additional relation this
comprises the most general system compatible with the three
assumptions above
$$
\eqalign{\r1 \L{n}1(\l) \r2 \L{n}2(\m) &= \L{n}2(\m) \rt2 \L{n}1(\l)
\bigl \lbrack (\rt3)^{-1} \r1 \r3 \bigr \rbrack , \cr
&\phantom{xxxxxxxxxxxxxxxxxxxxx}n = 1,\ldots,N \cr
(\r2)^{-1} \L{n}1(\l) \r2 \L{n+1}2(\m) &= \L{n+1}2(\m) (\r3)^{-1}
\L{n}1(\l) \r2 , \cr
&\phantom{xxxxxxxxxxxxxxxxxxxxx}n = 1,\ldots,N-1 \cr
(\r2)^{-1} \L{n}1(\l) \r2 \L{m}2(\m) &= \L{m}2(\m) (\r2)^{-1}
\L{n}1(\l) \r2 , \cr
&\phantom{xxxxxxxxxxxxxxxxxxxxx}1 \leq n < m \leq N,\ m \not= n + 1 .}
\eqn \refin
$$
This result corresponds to \refmark{\rHlKu} although
Yang-Baxterization of their constant algebra did not yield spectral
parameter dependent $R^2,R^3$. If we rewrite these equations
into one by attaching site labels to the $R$-matrices it is also
reminiscent of an equation in \Ref\rAFS{\AFS}, applied to the case of
non-ultralocal lattice current algebras (with constant $R$-matrices).

{\bf The monodromy.} By construction the monodromy $T(\l)$ defined as
in \mon\ satisfies
$$
\r1 T_1(\l) \r2 T_2(\m) = T_2(\m) \rt2 T_1(\l)
\bigl \lbrack (\rt3)^{-1} \r1 \r3 \bigr \rbrack , \eqn \genmo
$$
and in addition we can define the partial monodromy $T^{k,l}(\l) =
L^k(\l) L^{k-1}(\l) \cdots L^l(\l),\ k>l$ which has commutation
relations given by
$$
\eqalign{(\r2)^{-1} T^{k,l}_1(\l) \r2 T^{m,n}_2(\m) =  T^{m,n}_2(\m)
(R^3_{(k,n)}(\l,\m))^{-1} &T^{k,l}_1(\l) \r2 , \cr
&m>n>k>l .}
\eqn \pmon
$$
Here \pmon\ gives two equations with $R^3_{(k,k+1)} = R^3$, and
$R^3_{(k,n)} = R^2$ for $n \not= k+1$.

Readers will have noticed already that above formulas contain
fundamental commutation relations \fcr\ and equation \mon\ for the
monodromy of conventional (ultralocal) QISM as the special case
$R^2 = R^3 = I$. However, non-ultralocal models with nearest neighbour
interactions like the Toda chain which require $R^2 = I$ are not
compatible with our assumption that the monodromy satisfies the same
equation as the $L^n$. We restrict the discussion to completely
non-ultralocal cases with $R^2 \not= I, \ R^3 \not= I$.

Unlike in the conventional case here it is not straightforward to
obtain the set of commuting operators, the trace of \genmo\ over
auxiliary spaces does not factorize. The way out is to use a trick
developed in \refmark{\rSkl} which was generalized to monodromy
algebras of the type \genmo\ with four arbitrary $R$-matrices in
\refmark{\rFrMa}. In our case the theorem states that a solution $M$
of the following RE
$$
\eqalign{((\r1)^{t_1t_2})^{-1} M_1(\l) &(((\r2)^{t_1})^{-1})^{t_2}
M_2(\m) = \cr
&= M_2(\m) (((\rt2)^{t_2})^{-1})^{t_1} M_1(\l) ((\r4)^{t_1t_2})^{-1},}
\eqn \conre
$$
which further must satisfy $T_1(\l) M_2(\m) = M_2(\m) T_1(\l)$,
defines the desired operator
$$
t(\l) = \tr \lbrack M^t(\l) T(\l) \rbrack \eqn \mmon
$$
that obeys \zero. Superscripts $t$ (resp.\ $t_1,t_2$) indicate
transposition of the matrices (in auxiliary spaces), and we have put
$\r4 \equiv (\rt3)^{-1} \r1 \r3$. From \conre\ we can read off the
conditions such that $M=I$ is a solution, sufficient are
$$
((\r2)^{t_1})^{-1} = ((\r2)^{-1})^{t_1}, \qquad
((\r2)^{t_1t_2})^{-1} = ((\r2)^{-1})^{t_1t_2} . \eqn \mone
$$

{\bf Consistency conditions.} Several conditions must be imposed on
the $R$-matrices in \genmo\ such that the monodromy matrix has a
consistent quadratic algebra which can be used in the algebraic Bethe
ansatz, for example. They were studied in \refmark{\rFrMa}, in our
case they read
$$
\eqalign{R^1_{12}(\l,\m) R^1_{13}(\l,\nu) R^1_{23}(\m,\nu) &=
R^1_{23}(\m,\nu) R^1_{13}(\l,\nu) R^1_{12}(\l,\m) \cr
R^1_{12}(\l,\m) R^2_{31}(\nu,\l) R^2_{32}(\nu,\m) &=
R^2_{32}(\nu,\m) R^2_{31}(\nu,\l) R^1_{12}(\l,\m) \cr
R^4_{12}(\l,\m) R^2_{13}(\l,\nu) R^2_{23}(\m,\nu) &=
R^2_{23}(\m,\nu) R^2_{13}(\l,\nu) R^4_{12}(\l,\m) \cr
R^4_{12}(\l,\m) R^4_{13}(\l,\nu) R^4_{23}(\m,\nu) &=
R^4_{23}(\m,\nu) R^4_{13}(\l,\nu) R^4_{12}(\l,\m) .}  \eqn \cons
$$
They ensure also consistency of the first equation of \refin\ but as
for the other two extra conditions have to be imposed (which guarantee
that the braiding of the coproduct, in the sense of \refmark{\rMad},
embodied by them is indeed a braid group representation). If $R^3 =
R^2$ then there is only one extra condition, namely the YBE for $R^2$
$$
R^2_{12}(\l,\m) R^2_{13}(\l,\nu) R^2_{23}(\m,\nu) =
R^2_{23}(\m,\nu) R^2_{13}(\l,\nu) R^2_{12}(\l,\m) . \eqn \twocon
$$
In the general case the following YBE like conditions involving $R^3$
have to be added
$$
\eqalign{R^1_{12}(\l,\m) R^3_{31}(\nu,\l) R^3_{32}(\nu,\m) &=
R^3_{32}(\nu,\m) R^3_{31}(\nu,\l) R^1_{12}(\l,\m) \cr
R^4_{12}(\l,\m) R^3_{13}(\l,\nu) R^3_{23}(\m,\nu) &=
R^3_{23}(\m,\nu) R^3_{13}(\l,\nu) R^4_{12}(\l,\m) \cr
R^3_{12}(\l,\m) R^2_{13}(\l,\nu) R^2_{23}(\m,\nu) &=
R^2_{23}(\m,\nu) R^2_{13}(\l,\nu) R^3_{12}(\l,\m) \cr
R^2_{12}(\l,\m) R^2_{13}(\l,\nu) R^3_{23}(\m,\nu) &=
R^3_{23}(\m,\nu) R^2_{13}(\l,\nu) R^2_{12}(\l,\m) .}  \eqn \thrcon
$$
As can be expected they give no new conditions for $R^2$ if $R^3 =
R^2$. The example shows that not all three $R$-matrices neccessarily
have to be spectral parameter dependent. Anyway, the case $R^3 \not=
R^2$ in completely non-ultralocal models seems to be a less likely
possibility.

It can then be seen how the example of the introduction fits into this
formalism, namely $\r1 = S(\l,\m),\  R^2 = R^3 = \Rt$ (indeed
$(\Rt^3)^{-1} \r1 R^3 = \St(\l,\m)$), and $M = {\rm diag}(q^{-1},q)$
is a solution of \conre\ with this choice of $R$-matrices satisfying
\cons\ and \twocon.

{\bf Periodic case.} If we identify $n + N \equiv n$ two more
conditions have to be imposed on the $R$-matrices. The commutation
relation between $L^N$ and $L^1$ which are now on neighbouring sites
is described by the third equation of \refin\ instead of the second
one. This introduces an aperiodicity into the chain, in order to avoid
this it is neccessary to demand
$$
\r2 = (\rt2)^{-1}, \qquad \r3 = \r2 . \eqn \cona
$$

{\bf The case ${\bf R^2 = R^3 = \Rt^1}$.}
This particular choice of $R^2,R^3$ has two interesting properties.
Namely, we can ask whether it is possible to introduce a QG comodule
structure into \refin\ such that $L^n_T(\l) = T^{(n)}(\l) L^n(\l)
(T^{(n)}(\l))^{-1}$ also satisfies \refin. The answer is affirmative
if $\r2 = \r3 = \rt1$ and $T^{(n)}(\l)$ obeys extended QG relations
$$
\eqalign{\r1 T^{(n)}_1(\l) T^{(n)}_2(\m) &= T^{(n)}_2(\m)
T^{(n)}_1(\l) \r1 \cr
\r1 T^{(m)}_1(\l) T^{(n)}_2(\m) &= T^{(n)}_2(\m) T^{(m)}_1(\l) \r1 ,
\qquad m>n} \eqn \exqg
$$
and moreover $\L{n}1(\l) T^{(m)}_2(\m) = T^{(m)}_2(\m) \L{n}1(\l)$ for
all $m,n$. This is of course the spectral parameter dependent
counterpart of the QG comodule property of \re, but it is restricted
to this special choice of $R^2,R^3$ and does not hold for \refin\ in
general.

The second property in this case is that given certain ultralocal
algebras they can be used to realize \refin\ in terms of their
generators (the converse is not true of course). For example, an
algebra with generators $M^{\pm}$ and relations
$$
R(\l,\m) M^{\e_1}_1(\l) M^{\e_2}_2(\m) = M^{\e_2}_2(\m)
M^{\e_1}_1(\l) R(\l,\m) , \qquad (\e_1,\e_2) \in
\{(+,+),(+,-),(-,-)\} \eqn \yand
$$
allows to construct
$$
L^n(\l) = \S-1(\l) \cdots \S-{n-1}(\l) M^{\phantom{;}}_n(\l)
M^-_{n-1}(\l) \cdots M^-_1(\l) , \qquad n = 1,\ldots,N \eqn \local
$$
where $M_n(\l) = \S-n(\l) M^+_n(\l),\  \S-n(\l) \equiv
(M^-_n(\l))^{-1}$ and the $M^{\pm}_i(\l)$ are defined by a tensor
product analogous to \lipm. Then $L^n(\l)$ satisfies \refin\ with
$R(\l,\m) \equiv R^2(\l,\m) = R^3(\l,\m) = \Rt^1(\m,\l)$, and the
monodromy matrix \mon\ which obeys \genmo\ is found to be
$$
T(\l) = \S-1(\l) \cdots \S-N M^+_N(\l) \cdots M^+_1(\l) . \eqn \locmo
$$
However, this monodromy can be equivalently obtained by observing that
\yand\ implies the same commutation relations \yand\ for the
monodromies $ T^{\pm}(\l) = M^{\pm}_N(\l) \cdots M^{\pm}_1(\l)$, and
defining then $T(\l) = S(T^-(\l)) T^+(\l)$ we get \locmo. Whether
there are any advantages in constructing \refin\ out of such a local
algebra seems then questionable. An example where this construction
works is the Yangian double \refmark{\rDri} obeying \yand\ with
$R(\l-\m) = (\l-\m) I + h P$, and $h$ is the deformation parameter.
But for this case solutions of \conre\ seem not to exist.

{\bf Classical limit.} Finally, we conclude this section by mentioning
that \refin\ has a well defined semi-classical limit. Given the limit
$\r{} = I + i h r(\l,\m) + O(h^2)$, with a skew-symmetric classical
$r$-matrix $r_{21}(\m,\l) = - r_{12}(\l,\m)$, for all $R$-matrices of
\refin\ in case of infinitesimal quantization parameter $h$ and the
correspondence ${i \over h}\lbrack \> , \rbrack = \{\> , \}$ then the
following Poisson brackets are obtained from \refin
$$
\eqalign{\{\L{n}1(\l),\L{n}2(\m)\} &= r^1(\l,\m) \L{n}1(\l) \L{n}2(\m)
- \L{n}2(\m) \L{n}1(\l) \bigl \lbrack r^1(\l,\m) + r^3(\l,\m) -
{\widetilde r}^{\>3}(\m,\l) \bigr \rbrack \cr
&\phantom{=\ }+ \L{n}1(\l) r^2(\l,\m) \L{n}2(\m) - \L{n}2(\m)
{\widetilde r}^{\>2}(\m,\l) \L{n}1(\l) , \qquad n = 1,\ldots,N \cr
\{\L{n}1(\l),\L{m}2(\m)\} &= -r^2(\l,\m) \L{n}1(\l) \L{m}2(\m)
- \L{m}2(\m) \L{n}1(\l) r^2(\l,\m) + \L{n}1(\l) r^2(\l,\m) \L{m}2(\m)
\cr
&\phantom{=\ }+ \L{m}2(\m) r^3_{(n,m)}(\l,\m) \L{n}1(\l) , \qquad m>n}
\eqn \recl
$$
where as in \pmon\ we define $r^3_{(n,n+1)} = r^3$, and $r^3_{(n,m)} =
r^2$ if $m \not= n+1$. They reduce to the well known classical limit
of \fcr\ if we put $r^2 = r^3 = I$.


\chapter{CONCLUSIONS}

We conclude with a few remarks about possible applications. It would
be highly desirable to find more physical examples fitting this
formalism. In \refmark{\rHlKu} it was argued that supersymmetric
integrable models exhibit completely non-ultralocal commutation
relations. In these cases the commutation relations of the Lax
operators contain additional factors accounting for the statistics of
entries of the $L^n$. These $\pm$ signs can be conveniently
accommodated by the constant matrices $R^2 = R^3$. In that sense the
model of section 1 can be viewed as their $q$-generalization.

Further, it would be interesting to investigate the continuum limit of
\refin\ with Lax operators ${\cal L}(x;\l)$ obtained from
$$
L^n(\l) =\ :\ \buildrel \longleftarrow \over {\rm exp} \Bigl
(\int^{x_{n+1}}_{x_n} {\cal L}(x;\l) dx \Bigr ): \ = 1 + \Delta {\cal
L}(x;\l) + O(\Delta^2) , \qquad \Delta = x_{n+1} - x_n \eqn \contl
$$
where the colons around the path ordered exponential indicate normal
ordering which might be neccessary. One could expect in the
commutation relations of ${\cal L}(x;\l)$ beside $\delta(x-y)$ (and
possibly its derivative) also a nonlocal term like $\epsilon(x-y)$
from the condition $m>n$ in \refin\ which turns into $y>x$ in the
continuum limit.

Finally, there is a vague connection to 2+1 dimensions as exemplified
by the model of \refmark{\rAle} related to Chern-Simons theory. The
$L^n$ should then already be interpreted as monodromies along one
space direction, however, not commuting ones in contrast to the
situation in 2-dimensional statistical mechanics models.

\vskip2cm

{\bf Acknowledgements.} I should like to thank T. Miwa, N. Yu.
Reshetikhin and E. K. Sklyanin for discussion.

{\bf Note added.} When this work was completed I found a new paper by
L. Hlavaty where the same topic is discussed \Ref\rHla{\Hla}.

\refout
\vfill
\eject
\bye